\documentclass[12pt,twoside]{article}   
\usepackage[super,sort,comma]{natbib}

\usepackage{fancyhdr}		

\usepackage{amsmath}

\newcommand{\captionv}[3]{\begin{center}\parbox{#1cm}{\caption[#2]{{\sf #3}}}
        \end{center}}

\usepackage[section]{placeins}   %

\usepackage{graphicx}

\usepackage{booktabs}

\makeatletter \renewcommand\@biblabel[1]{$^{#1}$} \makeatother
\newcommand{\cen}[1]{\begin{center} #1 \end{center}}

\usepackage[mathlines]{lineno}

\usepackage{hyperref}
\hypersetup{ colorlinks,
    citecolor=blue,
    filecolor=blue,
    linkcolor=blue,
    urlcolor=blue
}

\usepackage{xcolor}

\definecolor{gray}{rgb}{0.6,0.6,0.6}
\definecolor{red}{rgb}{0.85,0,0}
\definecolor{green}{rgb}{0,0.85,0}
\definecolor{blue}{rgb}{0,0,0.85}
\definecolor{beige}{rgb}{0.92,0.87,0.78}
\usepackage[all]{hypcap}    

\usepackage{caption}

\begin{document}

\cen{\sf {\Large {\bfseries A Beam's Eye View to Fluence Maps 3D Network for Ultra Fast VMAT Radiotherapy Planning } \\  
\vspace*{10mm}
Simon Arberet\textsuperscript{1}, 
Florin C. Ghesu\textsuperscript{2},
Riqiang Gao\textsuperscript{1},
Martin Kraus\textsuperscript{2},
Jonathan Sackett\textsuperscript{3},
Esa Kuusela\textsuperscript{3},
Ali Kamen\textsuperscript{1}
} \\
\textsuperscript{1}Digital Technology and Innovation, Siemens Healthineers, Princeton, NJ, USA\\
\textsuperscript{2}Digital Technology and Innovation, Siemens Healthineers, Erlangen, Germany\\
\textsuperscript{3}Varian Medical Systems, a Siemens Healthineers Company, Helsinki, Finland
}

\pagenumbering{roman}
\setcounter{page}{1}
\pagestyle{plain}
simon.arberet@siemens-healthineers.com \\

\begin{abstract}
\noindent {\bf Background:} Volumetric Modulated Arc Therapy (VMAT) revolutionizes cancer treatment by precisely delivering radiation while sparing healthy tissues. 
Fluence maps generation, crucial in VMAT planning, traditionally involves complex and iterative, and thus time consuming processes. 
These fluence maps are subsequently leveraged for leaf-sequence. 
The deep-learning approach presented in this article aims to expedite this by directly predicting fluence maps from patient data.\\
{\bf Purpose:} To accelerate VMAT treatment planning by quickly predicting fluence maps from a 3D dose map. The predicted fluence maps can be quickly leaf sequenced because the network was trained to take into account the machine constraints.\\
{\bf Methods:} We developed a 3D network which we trained in a supervised way using a combination of $L^1$ and $L^2$ losses, and RT plans generated by Eclipse and from the REQUITE dataset, taking the RT dose map as input and the fluence maps computed from the corresponding RT plans as target.
Our network predicts jointly the 180 fluence maps corresponding to the 180 control points (CP) of single arc VMAT plans. In order to help the network, we pre-process the input dose by computing  the projections of the 3D dose map to the beam's eye view (BEV) of the 180 CPs, in the same coordinate system as the fluence maps. 
We generated over 2000 VMAT plans using Eclipse to scale up the dataset size. Additionally, we evaluated various network architectures and analyzed the impact of increasing the dataset size.
\\
{\bf Results:}
We are measuring the performance in the 2D fluence maps domain using  image metrics (PSNR, SSIM), as well as in the 3D dose domain using the dose-volume histogram (DVH) on a validation dataset.
The network inference, which does not include the data loading and processing, is less than 20ms. Using our proposed 3D network architecture as well as increasing the dataset size using Eclipse improved the fluence map reconstruction performance by approximately 8 dB in PSNR compared to a U-Net architecture trained on the original REQUITE dataset.
The resulting DVHs are very close to the one of the input target dose.
 \\
{\bf Conclusions:} 
We developed a novel deep learning approach for ultra fast VMAT planning by predicting all the fluence maps of a VMAT arc in one single network inference.
The small difference of the DVH validate this approach for ultra-fast VMAT planning.
 \\
\end{abstract}

\newpage     



\newpage

\pagenumbering{arabic}
\setcounter{page}{1}
\pagestyle{fancy}
\section{Introduction}

Volumetric modulated arc therapy (VMAT) is a modern radiation therapy technique that allows for the delivery of highly conformal dose distributions to tumors. 
One of the key steps in VMAT planning is the generation of fluence maps, which are used to control the intensity of the radiation beam as it rotates around the patient.

Traditionally, fluence maps for VMAT planning are generated within an optimization algorithm \cite{otto2008volumetric, liu2018comparison}. 
These algorithms can be computationally expensive and time-consuming, especially for complex cases. 
Indeed they need to perform many iterations between a leaf-sequencing step which tries to reproduce a target fluence map, and a gradient step which tries to minimize a dose objective,
and also rely on a multi-resolution approach which divide the arc in multiple sectors in order to avoid converging into local minima.

On the other hand, recent developments in predicting 3D radiation dose distributions from planning structure sets have shown promise in the literature \cite{mcintosh2017fully, kearney2018dosenet, zhan2022multi, gao2023flexible}. In this article, we leverages this progress, offering a modular framework for fluence map prediction from 3D dose maps. This modularization aims to simplify the radiation therapy planning process and improve prediction accuracy.

Recent articles have explored the application of deep learning for direct fluence map prediction in Intensity-Modulated Radiation Therapy (IMRT). However, existing methods \cite{wang2020fluence, lee2019fluence, wang2021deep} adopt an approach that processes each fluence map independently, inadvertently preventing the maps from synergistically contributing to the global treatment objective. This approach neglects the potential benefits of leveraging the complementarity between fluence maps. 

For VMAT, it's even more crucial to process all fields together because the continuous motion of the MLC requires an interconnected prediction strategy. Independent processing of individual fluence maps misses the necessary continuity of the MLC's dynamic leaf sequence.
Fluence map estimation for VMAT has also been explored in other studies \cite{ma2020deep, ma2021generalizability, zhu20223d}, where the authors used a 3D U-Net network to predict the fluence maps. 

In this article, we introduce a novel deep learning approach for predicting fluence maps in VMAT planning. Our method involves an initial transformation of the input 3D dose map into a new 3D representation, where the gantry's rotation corresponds to a translation that can be effectively utilized by CNNs. This transformation is achieved by projecting the dose map into the beam eye views (BEV) of each control point. Subsequently, we employ a 3D convolutional neural network architecture featuring advanced ConvNeXt \cite{liu2022convnet} building blocks to predict all fluence maps simultaneously.
The advantage of having a 3D CNN compared to a 2D CNN is that the convolution in the depth dimension which corresponds to the gantry rotation dimension is also processed with convolutions and the dynamic of the leaf-pairs motions and other gantry rotation equivariance behaviours can be well captured by convolutions thanks to their translation equivariance property.
On the other hand the final dose is the result of the contribution of all the fluence maps without any particular order, and as such long range dependencies along the gantry dimension are also important and it is where ConvNeXt blocks that mimic Transformers operations can help compared to more classical convolution blocks used in U-Net \cite{cciccek20163d} or ResNets blocks \cite{he2016deep}.   


Moreover we scaled up the training dataset, using Eclipse Scripting API (ESAPI) to recompute the optimization of plans of the REQUITE dataset \cite{seibold2019requite} for Varian VMAT single arc. 
This allows us to scale our dataset focused on Prostate VMAT single arc from about 100 cases to more than 2000 cases.
These plans are composed of 180 control points spaced by 2 degrees between each consecutive control point and doing a full arc rotation. 


This focus on a particular machine type is motivated by the interplay between machine constraints, Multi-Leaf Collimators (MLCs), and optimization algorithms, all of which are intrinsically machine-dependent.






\section{Methods}

In order to train a network to predict the fluence maps from a 3D dose map, we are using a dataset of RT plans which contains the optimized MLC positions and MU values of each of the control points covering the single VMAT arc as well as the 3D dose map calculated from these optimized MLC positions and MU values.
Using the MLC positions and MU values, we compute the fluence maps. Our fluence maps calculation model includes the leaf leakage and motion in between control points but ignores tongue-and-groove effect.
In order to simplify the task of the network, we transform the 3D dose map in the BEV of each control point, such that these dose projections are in the same geometry as the fluence maps.
Also as in this new representation, rotation of the gantry is transformed into a translation, we are using for the first time a 3D convolutional network architecture, instead of 2D as usually done in the litterature \cite{ma2020deep, ma2021generalizability}, in order to exploit translation equivariance in the gantry direction.

\subsubsection{Data}
We are using the REQUITE dataset and focus on Varian single arc VMAT plans.
The original REQUITE dataset \cite{seibold2019requite} contains 117 plans (which we split in 96 for training, 11 for validation), all prostate cases, with arcs covering a range of $358^\circ$ (almost full rotation) in 178 control points. In this collection of RT Plans, collimators angle varies depending of the plan but most of the cases have $30^\circ$ or $45^\circ$ angles and stay fix during each arc. The couch angle is zero for all the cases.
Among these 117 cases, 69 cases are using the Varian High-Definition 120 MLC (HD120) model from Varian, and 48 are using the Varian Millennium 120 ( M120) model. Both of these models are using 60 leaf pairs, but the HD120 model have a smaller leaf width (2.5 mm for the 32 inner leaf pairs and 5 mm for the 28 outer leaf pairs, vs 5 mm for the 40 inner leaf pairs and 10 mm for the 20 outer leaf pairs for the M120 model).
Most of the arc rotation directions are clock-wise and only 6 follow a counter clock-wise direction. Anyway, we arrange the 178 control points in increasing degree order (counter clock-wise) so that the network is agnostic to the rotation direction.
The plan can be executed in the other direction by simply reversing the order of the control points.

As this dataset is very small, one of our contribution is to increase the dataset size by using all the prostate plans of the REQUITE dataset, including those which were not planed for Varian single arc VMAT, and replan them using Eclipse Scripting API (ESAPI) \cite{gao2025automating}. 
This way we could increase the dataset size from 117 to around 2266 plans.
These new plans generated with Eclipse contain 180 control points equispaced by $2^\circ$, a fixed collimator angle of $30^\circ$. The couch angle is zero for all the cases.
Among these 2266 generated plans, 1161 cases are using the HD120 MLC, and 1105 are with the M120 MLC.

\subsubsection{Preprocessing}
\begin{figure}[ht]
   \begin{center}
   \includegraphics[width=14cm]{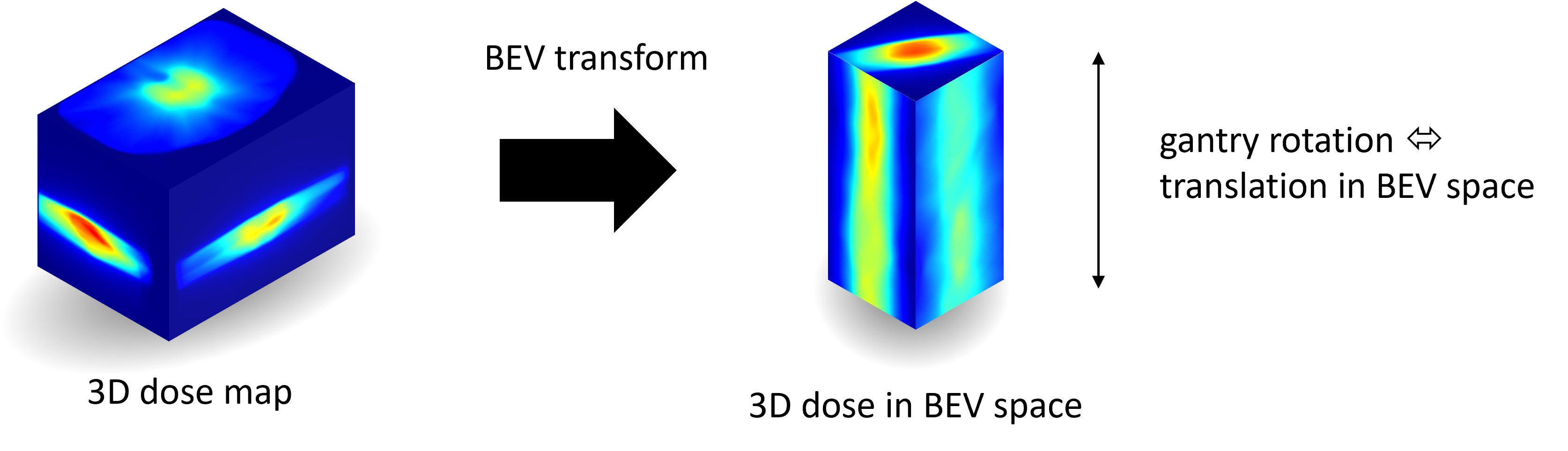}
   %
   %
   \captionv{16}{}{BEV transform of the 3D dose map.
   \label{fig_BEVtransform} 
    }  
    \end{center}
\end{figure}

We compute the fluence maps of the 180 control points of the single VMAT arc, as well as their corresponding dose BEV projections which are stacked to create a 3D tensor that is passed to our 3D network as illustrated in figure \ref{fig_BEVtransform}. We also compare our 3D network, which we present in the next section, with a 3D U-Net, and a 2D U-Net where the BEV projections are passed to the network in the input channel dimension. 
The calculation of these BEV projections take into account the gantry angle, the collimator angle, the couch angle (which was zero in all our experiments so actually had no effect in the projection), the isocenter, and the spacing and origins of the fluence maps and 3D dose maps respectively.

\subsubsection{Network architecture}

The 180 dose BEV projections are stacked to create a 3D tensor which is fed to the 3D network. These 180 BEV projections are sorted in increasing gantry angle degrees so that the network is agnostic to the gantry rotation direction.
The network predicts the corresponding 2D fluence maps on the 180 slices of the network output 3D volume. 

\begin{figure}[ht]
   \begin{center}
   \vspace{20pt} 
   \includegraphics[width=14cm]{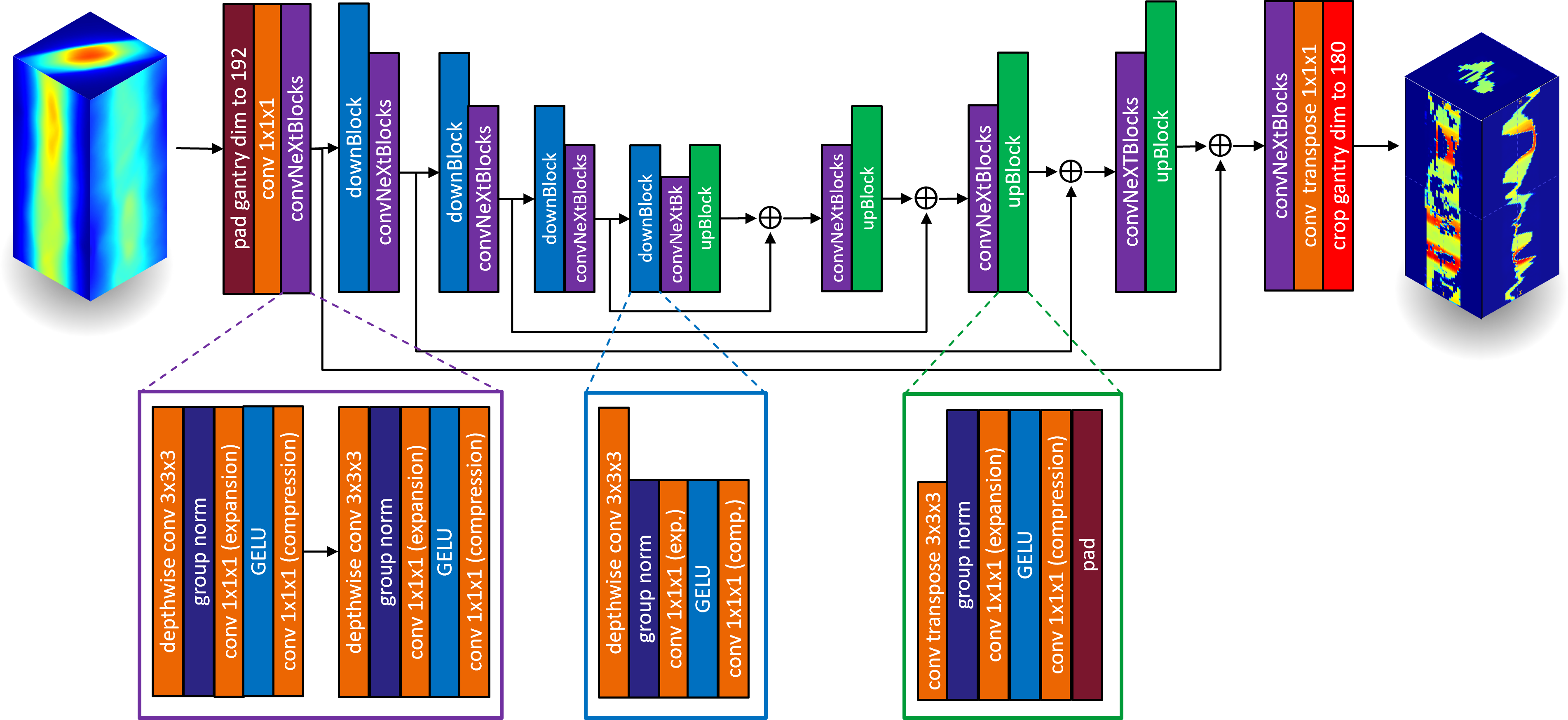}
   %
   %
   \captionv{16}{3D convolution network for VMAT fluence maps prediction}{
Our 3D network (3D MedNeXt) takes the BEV transform of the dose map as input, and output a 3D tensor where each slice is the fluence map prediction of one of the 180 control points.
   \label{fig_MedNeXT3D} 
    }  
    \end{center}
\end{figure}

Our network architecture, depicted in figure \ref{fig_MedNeXT3D}, has four downsampling steps which decrease the three dimensions of the feature maps by a factor 2. In order to accommodate for these 4 downsampling steps by a factor 2, we first perform a circular padding in order to increase the gantry dimension from 180 to 192, and then
the output of the network is cropped accordingly to reduce the gantry dimension back to 180.  


The network architecture is a 3D MedNeXt architecture \cite{roy2023mednext}, which is a 3D U-Net like encoder-decoder architecture, but where residual ConvNeXT blocks \cite{liu2022convnet} are used in place of the traditional convolutional blocks, up and downsampling blocks.
ConvNeXt blocks were introduced in order to create a simple, efficient and scalable architecture that combines the strengths of both CNNs and Transformers and showed improved performance compared to ViT \cite{dosovitskiy2020image} and Swin Transformer \cite{liu2021swin}.
Each ConvNeXt block in our architecture contains a depthwise convolution layer with a kernel size of $3\times3\times3$ followed by a channel-wise group norm, then a $1\times1\times1$ convolution (expansion layer) which expands the number of channels by a factor 4 , 
followed by a Gaussian Error Linear Unit (GELU) activation function, followed by a $1\times1\times1$ convolutional layer (compression layer) which performs a channel-wise compression of the feature maps by a factor 4.
These three layers (depthwise convolution, expansion and compression) mimic the Transformer block architecture.





We compare this 3D MedNeXt model with a 2D U-Net \cite{ronneberger2015u} that stacks the 180 control points in the input and output channel dimensions and a 3D U-Net \cite{cciccek20163d} that perform the same data preprocessing (BEV projections, padding) and post-processing (cropping) as our 3D MedNeXt.
These 2D and 3D U-Nets contain 4 downsampling and upsampling layers (i.e. 5 scales) as for our 3D MedNeXt network, and 64 channels in the first convolutional layer and the number of channels is doubled after each downsampling layer.



\subsubsection{Experiments}

We trained our networks in a supervised way, using a combination of L1 and L2 loss, ADAM optimizer with a learning rate of 0.0001 and a batch size of 1.

Our dataset of 2266 plans was split in 1868 plans for the training and 193 for the validation. 
We used Peak signal-to-noise ratio (PSNR)\footnote{The maximum value of the target fluence maps is used as the “peak” value in the PSNR equation.} and structural similarity index measure (SSIM) as an image-based metrics to measure the error between the network predicted fluence maps and the target fluence maps.
We used these two metrics to compare the methods in our ablation study and select the best method. Then we computed the dose of the best method using the Acuros AXB dose calculation function in order to compute the mean absolute error (MAE) in Gy, as well as generating the dose-volume histograms (DVH) of the target vs our network prediction.

In a first experiment we compare the 2D and 3D U-Nets which are commonly used in the litterature \cite{ma2020deep, ma2021generalizability, lee2019fluence, wang2020fluence}, with our 3D network on the original REQUITE dataset (Varian VMAT single arc), i.e. by training and testing on the original REQUITE data of 117 plans.

Then we train our models on the full set of Eclipse generated plans as well as two randomly selected subsets of this dataset (see table \ref{tab:datasets}): one of 117 plans as in the original REQUITE data, and another of 500 plans (411 for training, 44 for validation) and evaluate these models on both the full dataset and the subset datasets in order to assess the effect of the dataset size on the performance. 

Finally we compute the dose and compute the DVHs in order to validate that the predicted fluence maps of the best method can reproduce the target dose. 


\begin{table}
\centering
\caption{Dataset names and the number of plans they contains.}
\label{tab:datasets}
\vspace*{2ex}
\begin{tabular}{lcc}
\toprule
dataset name & short name &  Total \# of plans (\# in training / \# in validation) \\
\midrule
Eclipse REQUITE (full) & Ecl. full & 2266 (1868 / 193) \\
Eclipse REQUITE 500 & Ecl. 500 & 500 (411 / 44)  \\
Eclipse REQUITE 117 & Ecl. 117 & 117 (96 / 12)  \\
Original REQUITE & orig. REQ.& 117 (96 / 11)  \\
\bottomrule
\end{tabular}
\end{table}

\section{Results}

The quantitative results on the original REQUITE dataset are depicted in table \ref{tab:comparison_REQUITE}.
As can be observed, the propose 3D network improved PSNR and SSIM, 5.58 dB compared to 2D U-Net is and 3 dB compared to 3D U-Net.

\begin{figure}[ht]
   \begin{center}
   \vspace{20pt} 
   \includegraphics[width=14cm]{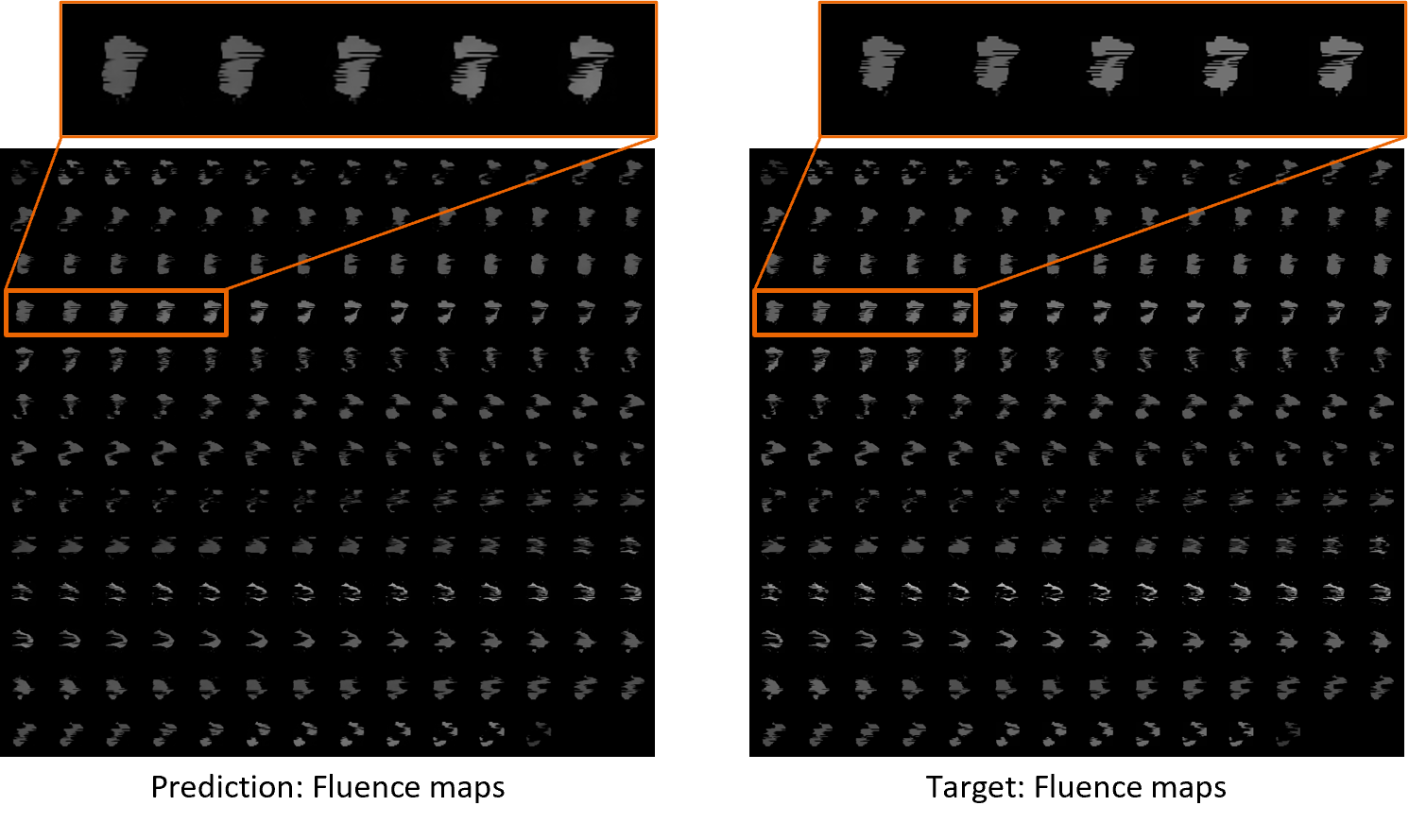}
   %
   %
   \captionv{16}{3D convolution network for VMAT fluence maps prediction}{
Example from the validation set, showing the fluence maps of the 180 control points of a VMAT plan predicted by our 3D MedNeXt network compared to the corresponding target fluence maps.
   \label{fig_fluence_maps} 
    }  
    \end{center}
\end{figure}

The quantitative results on the Eclipse generated dataset are depicted in table \ref{tab:comparison_Eclipse} and an example of fluence maps prediction is depicted in figure \ref{fig_fluence_maps}.
We can observe again that our 3D network improves significantly the PSNR and SSIM over the 2D and 3D U-Net, 5.55 dB compared to 2D U-Net is and 6.51 dB compared to 3D U-Net.
We can also notice that increasing the dataset size improved significantly the performance.
For our 3D MedNeXt, the increase from $\sim$100 (Ecl. 117) plans to $\sim$400 (Ecl. 500) plans in the training led to an increase of performance of 2.1 dB, and the increase from $\sim$400 plans (Ecl. 500) to $\sim$1900 plans (Ecl. full)
 led to an increase of performance of 2.69 dB. So the total increase of performance due to the dataset size increase from $\sim$100 (Ecl. 117) to $\sim$1900 plans (Ecl. full) was 4.79 dB.

\begin{table}
\centering
\caption{Comparison of our 3D MedNeXt with the 2D and 3D U-Net (baselines), all trained and validated on the original REQUITE dataset of 117 plans (96 for training, 11 for validation).}
\label{tab:comparison_REQUITE}
\vspace*{2ex}
\begin{tabular}{lc}
\toprule
Method &  PSNR/SSIM on  original REQUITE \\
\midrule
3D MedNeXt trained on original REQUITE & \textbf{24.91 dB} / \textbf{0.9325} \\
3D U-Net trained on original REQUITE & 21.91 dB / 0.3588  \\
2D U-Net trained on original REQUITE & 19.33 dB / 0.7657  \\
\bottomrule
\end{tabular}
\end{table}

\begin{table}
\centering
\caption{Comparison of our 3D MedNeXt network with the 2D and 3D U-Net (baselines), trained and validated on the Eclipse REQUITE dataset (2266 plans), and subsets of 500 plans and 117 plans.}
\label{tab:comparison_Eclipse}
\vspace*{2ex}
\begin{tabular}{lcc}
\toprule
Method &  PSNR/SSIM on Ecl. 500  & PSNR/SSIM on Ecl. full \\
\midrule
3D MedNeXt trained on Ecl. full & \textbf{28.72 dB} / \textbf{0.9721} & \textbf{28.71 dB} / \textbf{0.9712} \\
3D MedNeXt trained on Ecl. 500 & 26.01 dB / 0.9485 & 26.02 dB / 0.9481  \\
3D MedNeXt trained on Ecl. 117 & 23.90 dB / 0.9191 & 23.92 dB / 0.9199  \\
3D U-Net trained on Ecl. full & 21.90 dB / 0.2458 & 22.20 dB / 0.2664 \\
3D U-Net trained on Ecl. 500 & 21.85 dB / 0.5494 & 22.00 dB / 0.5618  \\
3D U-Net trained on Ecl. 117 & 17.76 dB / 0.1548 & 17.97 dB / 0.1687  \\
2D U-Net trained on Ecl. full  & 23.18 dB / 0.8918 & 23.16 dB / 0.8916 \\
2D U-Net trained on Ecl. 500  & 21.80 dB / 0.8543 & 21.82 dB / 0.8548 \\
2D U-Net trained on Ecl. 117  & 20.46 dB / 0.7845 & 20.64 dB / 0.7963 \\
\bottomrule
\end{tabular}
\end{table}

We also depicted in figure \ref{tab:dvh} few DVHs examples obtained by our 3D MedNeXt network. They are randomly selected from the validation set of the full Eclipse datasset.
We also inspected all the DVHs of the validation dataset, and fund that they all looked similar in accuracy as the one depicted in figure \ref{tab:dvh}.



\begin{table}
\centering
\captionsetup{position=bottom} 
\begin{tabular}{cc}  
  \includegraphics[width=0.45\textwidth]{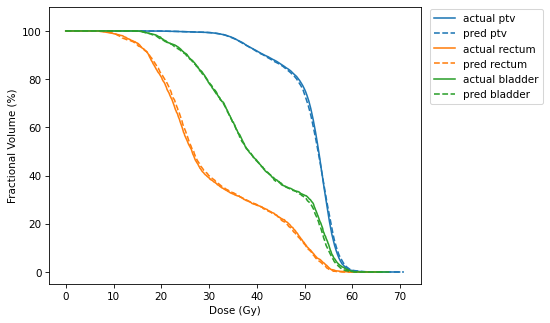} & \includegraphics[width=0.45\textwidth]{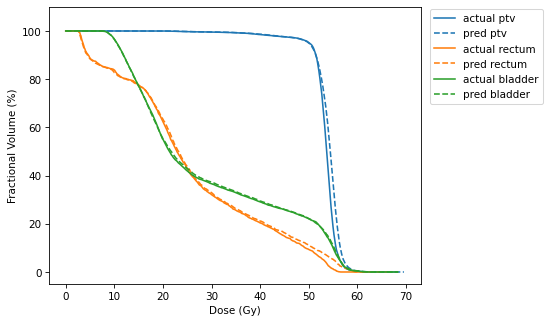} \\
  \includegraphics[width=0.45\textwidth]{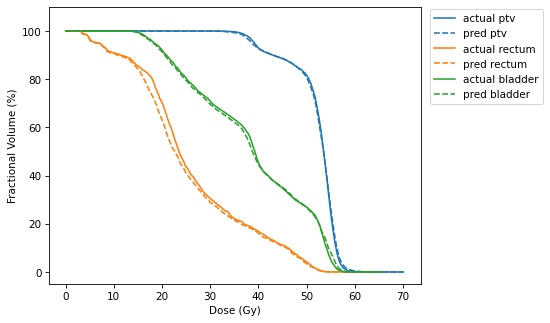} & \includegraphics[width=0.45\textwidth]{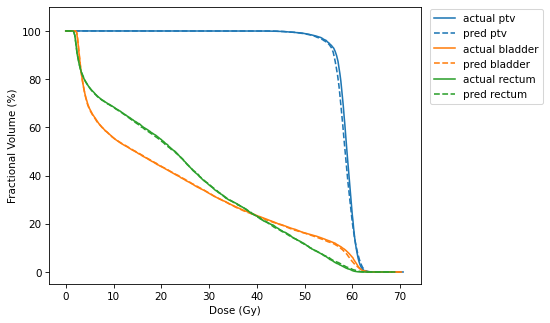} \\
\end{tabular}
\captionof{figure}{DVHs of our 3D MedNeXt network. We show four examples from our validation dataset. The plain lines are the DVHs calculated from the target fluence maps and the dashed lines are the DVHs calculated from the fluence maps predicted by our network. \label{tab:dvh}}
\end{table}

\section{Discussion}

The proposed approach introduces a 3D deep-learning network that predicts all the fluence maps of single arc VMAT plan with a single network inference, requiring less than 20ms.
Of course other modules are necessary to perform the inverse planning such as dataloading and preprocessing, dose prediction and leaf sequencing, so the total time for an inverse planning will be significantly longer.

The main idea of our method is to first performing a BEV transform of the input dose in order to transform the 3D dose map in a more practical representation.
This BEV representation has the advantage of 1) having the dose projection in the same geometry as the fluence maps in order to help the network, and 2) transforming the gantry rotation to a translation in the BEV space, which is convenient in order to take advantage of the convolutions that are the core component of CNNs.
For this reason we used a 3D CNN architecture as opposed to a 2D CNN, so that in addition to the two spatial dimensions of the fluence maps, the gantry motion/rotation is also processed with convolutions.
This local processing is accounting for all the local shape (spatial dimensions), and dynamics of the MLC (gantry dimension).
While the shape and dynamics of the MLC can be well captured by the local 3D convolutional kernels, there are also global features that need to be capture such as the fact that the total dose is the result of the contribution of all the control points.
For this reason, ConvNeXt blocks which implement fully connected layers in the feature dimension though 1x1x1 convolutions and mimics the architectures of Transformers allow such processing as opposed to classical U-Net architectures based on convolutional blocks or ResNet blocks.

Another important contribution of this paper was to create additional plans and show the importance of scaling-up the training dataset size for the overall performance of AI-based fluence prediction.
In particular, our ablation study showed an increase of performance of about 2.5 dB in PSNR each time the dataset size was multiplied by a factor 4.


This advancement offers promising prospects for ultra-fast inverse planning within an end-to-end framework, where dose computation \cite{gao2023flexible} precedes fluence map generation, followed by leaf-sequencing \cite{gaomulti}.
Additionally, it could serve as an initialization for established VMAT optimizers like the Photon Optimization Algorithm (PO) of Varian \cite{varian2015eclipse}.
It's noteworthy that, while this study focused on prostate plans, future endeavors aim to extend the methodology to encompass other anatomical regions such as lung and head and neck. We also envisage extending the method to accommodate multiple arcs VMAT and Hybrid IMRT/VMAT plans.  

In order to improve the fluence maps prediction, we considered inputting other informations to the network such as the CT scan, the organ contours, and in particular the PTV.
These data could be incorporated similar to the dose data, as an extra input channel through the computation of their BEV projections. 
While we explored this approach, our experiments did not reveal significant improvements.  
One possible explanation could be that a substantially larger dataset might be required to fully exploit the potential of the additional information. 



\section{Conclusion}

We have developed a novel AI-based method for fluence maps prediction of VMAT plans from 3D dose maps. Our method predicts all the fluence maps of a VMAT arc at once using a 3D network that takes as input the beam's eye view projections of the 3D dose map.
We also improved the performance of our fluence prediction by generating more than 2000 plans using Eclipse and showed the importance of scaling up the dataset size for improved performance.
Our network inference is very fast (less than 20ms) enabling ultra-fast inverse planning, while simultaneously enhancing PSNR by $24\%$ compared to a 2D U-Net trained on the same dataset and even $49\%$ compared to a 2D U-Net trained on the original REQUITE dataset. 
We studied variants of the network (2D and 3D U-Net vs 3D MedNeXt) as well as the effect of the dataset size. 
We think that our proposed 3D network can be used as a module within an ultra-fast inverse planning framework, or as an improved initialization method for an iterative VMAT optimizer.
Future research directions include extending the method to multiple arcs VMAT and augmenting the training dataset size to further enhance its efficacy and generalizability.

\section{Disclaimer}
The concepts and information presented in this paper / presentation are based on research results that are not commercially available. Future commercial availability cannot be guaranteed.

\section{Acknowledgments}
We acknowledge the use of the REQUITE dataset in this work. The REQUITE consortium includes Catharine West, Jenny Chang-Claude, Chris Talbot, Liv Veldeman, Dirk De Ruysscher, Barry Rosenstein, Tiziana Rancati, Ana Vega, Sara Gutiérrez-Enríquez, David Azria, Ananya Choudhury, Elena Sperk, Petra Seibold, Adam Webb, Erik Briers, Hilary Stobart, and Tim Ward. REQUITE received funding from the European Union's Seventh Framework Programme for research, technological development, and demonstration under grant agreement no. 601826. \\

\clearpage


\section*{References}
\addcontentsline{toc}{section}{\numberline{}References}
\vspace*{-20mm}





\bibliography{./bev2fluence_9_16_arxiv}      



\bibliographystyle{./medphy.bst}    


\end{document}